\documentclass[twocolumn,prl,aps,superscriptaddress]{revtex4}
\usepackage{graphicx}

\begin{document}
\title{Ba$_{1-x}$K$_x$Mn$_2$As$_2$: An Antiferromagnetic Local-Moment Metal}
\author{Abhishek Pandey}
\altaffiliation{apandey@amelab.gov} 
\author{R. S. Dhaka}
\author{J. Lamsal}
\author{Y. Lee}
\author{V. K. Anand}
\author{A. Kreyssig}
\affiliation {Ames Laboratory and Department of Physics and Astronomy, Iowa State University, Ames, Iowa 50011, USA}
\author{T. W. Heitmann}
\affiliation {The Missouri Research Reactor, University of Missouri, Columbia, Missouri 65211, USA}
\author{R. J. McQueeney}
\author{A. I. Goldman}
\author{B. N. Harmon}
\author{A. Kaminski}
\author{D. C. Johnston} 
\altaffiliation{johnston@ameslab.gov}
\affiliation {Ames Laboratory and Department of Physics and Astronomy, Iowa State University, Ames, Iowa 50011, USA}

\date{February 29, 2012}

\begin{abstract}

The compound $\rm{BaMn_2As_2}$ with the tetragonal $\rm{ThCr_2Si_2}$ structure is a local-moment  antiferromagnetic insulator with a N\'eel temperature $T_{\rm N}=625$~K and a large ordered moment \mbox{$\mu = 3.9~\mu_{\rm B}$}/Mn.  We demonstrate that this compound can be driven metallic by partial substitution of Ba by K, while retaining the same crystal and antiferromagnetic structures together with nearly the same high $T_{\rm N}$ and large~$\mu$.  Ba$_{1-x}$K$_x$Mn$_2$As$_2$ is thus the first metallic $\rm{ThCr_2Si_2}$-type $M$As-based system containing local $3d$ transition metal $M$ magnetic moments, with consequences for the ongoing debate about the local moment versus itinerant pictures of the FeAs-based superconductors and parent compounds.  The Ba$_{1-x}$K$_x$Mn$_2$As$_2$ class of compounds also forms a bridge between the layered iron pnictides and cuprates and may be useful to test theories of high $T_{\rm c}$ superconductivity.

\end{abstract}


\maketitle

Superconducting transition temperatures $T_{\rm c} > 50$~K have been observed for only two classes of materials---layered cuprates and iron arsenides \cite{Johnston1997, Johnston2010}. Both classes contain stacked square lattice layers of the transition metal atoms.  However, the parent compounds of the two families exhibit divergent physical properties.  For example, $\rm{La_2CuO_4}$ is a local magnetic moment antiferromagnetic (AF) insulator \cite{Johnston1997} while $\rm{BaFe_2As_2}$ is metallic and its AF ordering is widely considered to be best characterized as a spin-density wave arising from conduction carriers \cite{Johnston2010}. These differences create barriers for a general and comprehensive understanding of the underlying mechanisms of high-$T_{\rm c}$ superconductivity and related phenomena in a broad spectrum of materials.  Thus, it is desirable to create a material that can bridge the gap between the cuprates and iron arsenides.  Herein we report the synthesis and properties of such a material, Ba$_{1-x}$K$_x$Mn$_2$As$_2$ ($x = 0.016, 0.05$), which shares properties with both classes.

The undoped parent compound $\rm{BaMn_2As_2}$ crystallizes in the same body-centered-tetragonal (bct) $\rm{ThCr_2Si_2}$-type structure as the $M\rm{Fe_2As_2}$ ($M$ = Ca, Sr, Ba) iron arsenide parent compounds do at room temperature \cite{Johnston2010, Brechtel1978,Singh2009a}. It is a semiconductor with an activation energy of $\sim 30$~meV determined from electrical resistivity $\rho(T)$ measurements \cite{Singh2009a, An2009}, consistent with electronic structure calculations that indicate a band gap of $\sim 100$--150~meV \cite{An2009}.  Heat capacity $C_{\rm p}$ measurements at low-$T$ yield an electronic linear heat capacity coefficient $\gamma = 0$ which is  consistent with an insulating ground state \cite{Singh2009a}.  $\rm{BaMn_2As_2}$ orders into a G-type (N\'eel- or checkerboard-type) AF structure below a N\'eel temperature $T_{\rm N}$ = 625(1)~K with an ordered moment at 10~K of $\mu= 3.88(4)~\mu_{\rm B}$/Mn aligned along the crystallographic $c$~axis \cite{Singh2009a, Johnston2010, Singh2009b}.  Since $\rm{BaMn_2As_2}$ is an insulator at low temperatures, these results demonstrate that the antiferromagnetism arises from ordering of local Mn magnetic moments instead of from itinerant current carriers.  Both the static and dynamic magnetic properties for $T = 4$--1000~K are well-described by the AF $J_1$-$J_2$-$J_c$ local moment Heisenberg model, with a Mn spin $S = 5/2$ as expected from the $3d^5$ electronic configuration of Mn$^{+2}$ \cite{Johnston2011}.

We previously reported that many transition metals do not substitute for Mn in $\rm{BaMn_2As_2}$ crystals at levels above 0.5\% \cite{Pandey2011}.  Exceptions are Cr or Fe that substitute at levels of 4.4\% and $\lesssim 10$\%, respectively.  However, these two doped compounds show electronic transport and magnetic behaviors very similar to those of the undoped AF insulator $\rm{BaMn_2As_2}$ \cite{Pandey2011}.  We now report the successful doping of K for Ba to form the new system Ba$_{1-x}$K$_x$Mn$_2$As$_2$.  Our $\rho(T)$, $C_{\rm p}(T)$ and angle-resolved photoemission spectroscopy (ARPES) measurements demonstrate that the ground states of a single crystalline sample with  $x=0.016$ and a polycrystalline sample with $x=0.05$ are metallic.  Spin-polarized electronic structure calculations confirm that undoped $\rm{BaMn_2As_2}$ is a band insulator (semiconductor) whereas a hole band crosses the Fermi energy $E_{\rm F}$ centered at the $\Gamma$ point for $x = 0.016$ and 0.05.  On the other hand, our magnetic susceptibility $\chi$ and neutron diffraction (ND) measurements for $x = 0.016$ and 0.05 show the same local moment AF ordering behavior as in undoped  $\rm{BaMn_2As_2}$.  The ND measurements for $x=0.05$ exhibit a small reduction of $T_{\rm N}$ to 607(2)~K from 625~K for $x=0$ and a slight increase in $\mu$ to 4.21(6)~$\mu_{\rm B}$/Mn from 3.88(4)~$\mu_{\rm B}$/Mn.  The fact that $T_{\rm N}$ and $\mu$ of the K-doped sample are nearly identical to those of the undoped AF insulator $\rm{BaMn_2As_2}$ demonstrates that the magnetic ordering in the K-doped sample also arises from local moments.   Our results thus establish Ba$_{1-x}$K$_x$Mn$_2$As$_2$ in our doping range as a hole-doped AF local-moment metal.  This new system and obvious extensions open many interesting avenues for investigating both theoretically and experimentally  the interactions of itinerant carriers with local moments in MnAs-based compounds with the ${\rm ThCr_2Si_2}$ structure.  Of particular interest is the potential for high-$T_{\rm c}$ superconductivity \cite{Johnston2010, Singh2009a, Singh2009b}.

Polycrystalline samples with nominal compositions $\rm{KMn_2As_2}$, $\rm{BaMn_2As_2}$ and $\rm{Ba_{0.95}K_{0.05}Mn_2As_2}$ were synthesized by solid state reaction \cite{Pandey2011}, except that the $\rm{KMn_2As_2}$ and $\rm{Ba_{0.95}K_{0.05}Mn_2As_2}$ were sealed inside Ta tubes instead of placed in alumina crucibles.  The polycrystalline samples were used for crystal growths using Sn as solvent \cite{Pandey2011}. The chemical compositions of the crystals were determined by energy dispersive x-ray analysis, which yielded $x=0.016(6)$ for the K-doped crystals utilized here. No Sn from the flux was detected in the crystals.  The electronic transport, thermal and magnetic properties were measured using Quantum Design, Inc., measurement systems.  The ARPES data for crystals with $x=0$ and 0.016 were acquired using a laboratory-based system consisting of a Scienta SES2002 electron analyzer, GammaData UV lamp and custom-designed refocusing optics. ND measurements on a polycrystalline sample with $x=0.05$ were performed on the powder diffractometer at the University of Missouri Research Reactor using neutrons of wavelength $\lambda = 1.479$~\AA\@. Analysis of the ND data was performed by the Rietveld method using {\tt FULLPROF} \cite{Carvajal1998}. The spin-polarized electronic structures for $x$ = 0, 0.016 and 0.05 were calculated using a full-potential linear augmented plane wave method \cite{Blaha2001} with a local density approximation functional \cite{Perdew1992}.  The experimental structural data at 10~K were taken from Ref.~\cite{Singh2009b}.  To determine the influence of K-doping, we employed the virtual crystal approximation.

\begin{figure}
	\includegraphics[width=3in]{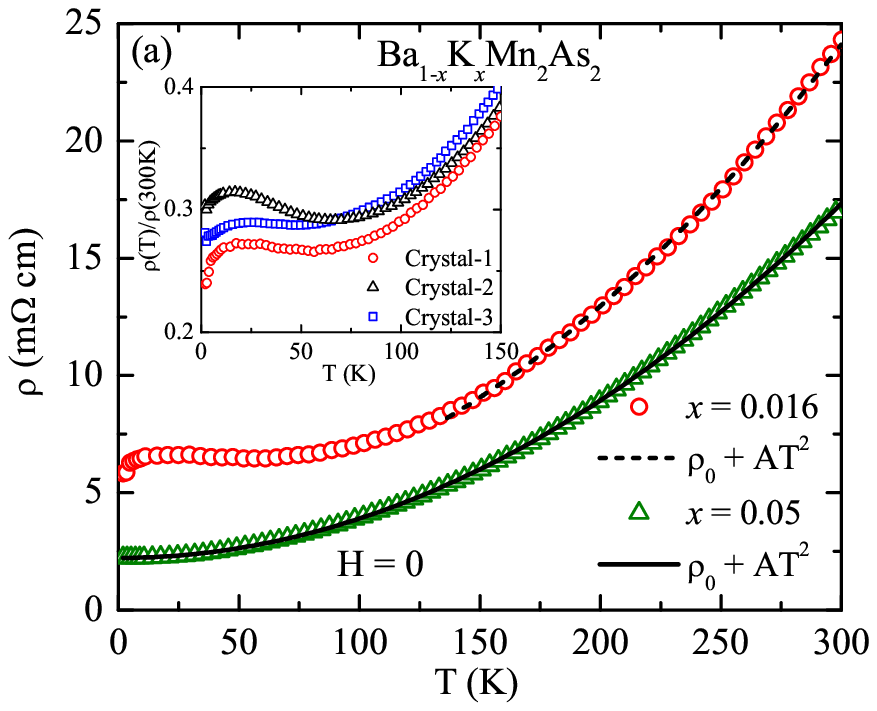}
		\includegraphics[width=2.9in]{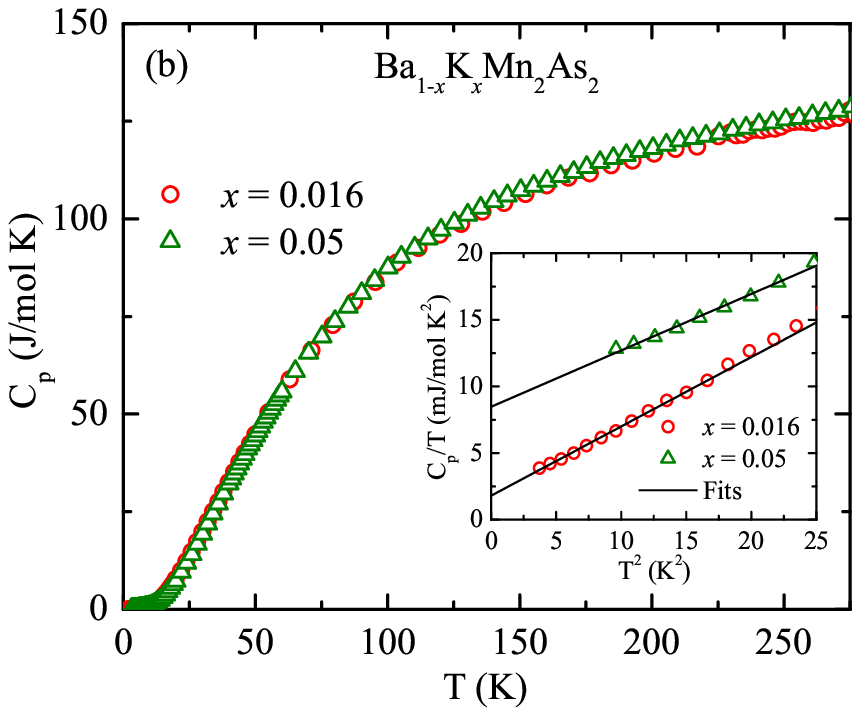}
\caption{(Color online) (a) Electrical resistivity $\rho$ versus temperature $T$ in the $ab$~plane of a crystal with $x=0.016$ and $\rho(T)$ for a polycrystalline sample with $x=0.05$. Fits by the expression $\rho = \rho_{0} + AT^2$ are shown by the black dashed and solid curves, respectively.  Inset: Expanded plots of $\rho(T)$ at low~$T$ for three crystals from the same batch.  (b)  Heat capacity $C_{\rm p}$ versus $T$ of the two samples in (a). Inset: $C_{\rm p}/T$ versus $T^2$ at $T<5$~K\@.}
	\label{fig:Figure_Resistivity}
\end{figure}

Electron correlation effects seem to be important in the metallic K-doped crystals as evidenced from our $\rho(T)$ and $C_{\rm p}(T)$ data.  Figure~\ref{fig:Figure_Resistivity}(a) shows $\rho(T)$ in the $ab$~plane for a crystal with $x=0.016$ and $\rho(T)$ for a polycrystalline sample with $x=0.05$. The $\rho(T)$ data for both samples suggest metallic ground states.  However, the magnitudes of $\rho \approx$ 5.8 and 2.2~m$\Omega$\,cm at 2~K for $x = 0.016$ and 0.05, respectively, are rather large compared to most metallic conductors. The large magnitudes of $\rho$ are evidently due to the low hole concentrations arising from the low K doping levels.  The $\rho(T)$ data in Fig.~\ref{fig:Figure_Resistivity}(a) indicate that $\rho$ decreases with increasing hole concentration $x$ as one would expect.  We fitted the $\rho(T)$ data for both samples by the expression $\rho(T) = \rho_{0} + A T^2$. While the data for $x = 0.05$ are well-described over the entire $T$ range of the measurement, the data for $x = 0.016$ deviate from the $T^2$-dependence below 135§~K and are therefore fitted above this $T$\@. The fitted values of $A$ are $A = 2.232(4)\times10^{-4}$ and $1.6807(6)\times10^{-4}$~m$\Omega$\,cm/K$^2$ for $x=0.016$ and 0.05, respectively. The $T^2$ behavior suggests that hole-hole scattering  may be mainly responsible for the $T$ dependence of $\rho$ \cite{Baber1937}. The low-$T$ in-plane $\rho(T)$ behaviors of three crystals from the same batch with $x \approx 0.016$ are shown in the inset in Fig.~\ref{fig:Figure_Resistivity}(a), which reproducibly show a leveling off or a small increase in $\rho$ below $\approx 70$~K, followed by a decrease below $\approx 15$~K\@.  The origins of these two low-$T$ behaviors require further investigation.

The $C_{\rm p}(T)$ data for the two samples in Fig.~\ref{fig:Figure_Resistivity}(a) are shown in Fig.~\ref{fig:Figure_Resistivity}(b). The values of $C_{\rm p}$ at $T=275$~K are slightly larger than the classical Dulong-Petit lattice heat capacity value at constant volume  $C_{\rm V} = 15R \approx124.7$~J/mol~K\@. The low-$T$ data in the inset in Fig.~\ref{fig:Figure_Resistivity}(b) were fitted by $C_{\rm p}/T = \gamma + \beta T^2$, yielding  $\gamma = 1.8(2)$ and 8.4(4)~mJ/mol~K$^2$ and $\beta = 0.52(3)$ and 0.43(4)~mJ/mol~K$^4$ for $x = 0.016$ and 0.05, respectively.  Consistent with the $\rho(T)$ data, the nonzero values of $\gamma$ indicate that the K-doped samples are metallic. The densities of states at $E_{\rm F}$ estimated from the $\gamma$ values \cite{Kittel2005} are 0.76 and 3.6~states/(eV\,f.u.)\ for both spin directions for $x=0.016$ and 0.05, respectively, where f.u.\ means formula unit.  The Debye temperatures obtained from  the $\beta$ values \cite{Kittel2005} are $\Theta_{\rm D} = 265(5)$ and 283(9)~K for $x = 0.016$ and 0.05, respectively.  The Kadowaki-Woods (KW)0 ratios $R_{\rm KW} = A/\gamma^2$  \cite{Kadowaki1986} are $R_{\rm KW}=69$ and 2.4~m$\Omega\,{\rm cm\,mol^2J^{-2}K^2}$ for $x = 0.016$ and 0.05, respectively. These values are 2--4 orders of magnitude larger than the value $R_{\rm KW} \sim 0.010~{\rm m}\Omega\,{\rm cm\,mol^2J^{-2}K^2}$ observed for a number of heavy fermion compounds \cite{Kadowaki1986}, but are much closer to the trend for the recently revised KW relation \cite{Jacko2009}. In either case, our combined $\rho(T)$ and $C_{\rm p}(T)$ data at low $T$ suggest the presence of strong electron correlation effects in the Ba$_{1-x}$K$_x$Mn$_2$As$_2$ system.

\begin{figure}
	\includegraphics[width=\columnwidth]{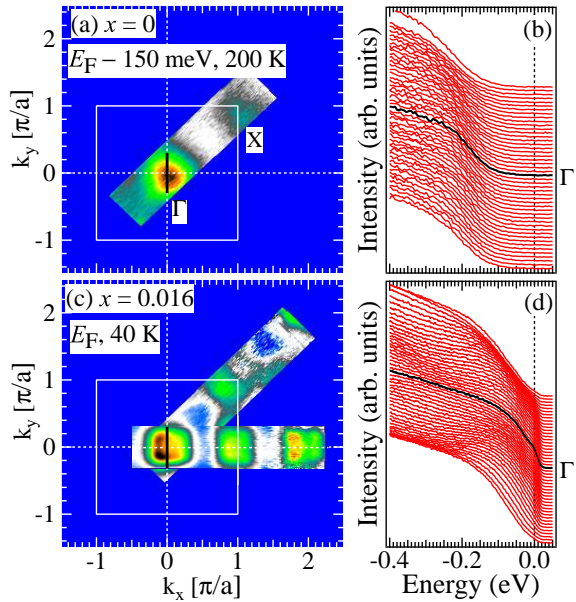}
	\caption{(Color online) ARPES intensity map at 150~meV below $E_{\rm F}$ (a) and EDCs (b) for BaMn$_2$As$_2$ at 200~K\@.  Also shown are the intensity map at $E_{\rm F}$ (c) and EDCs (d) for Ba$_{0.984}$K$_{0.016}$Mn$_2$As$_2$ at 40~K\@.  Each map in (a) and (c) is obtained by integrating over an energy window of $\pm$10~meV\@.  The paths of the EDCs in (b) and (d) are shown as vertical black lines through the $\Gamma$ points in (a) and (c), respectively.  The outline of the first Brillouin zone in (a) and (c) is shown by a white square.}
	\label{fig:Figure_ARPES}
\end{figure}

For the ARPES measurements, the chemical potential $E_{\rm F}$ was determined from measurements on a Au reference sample.  Since the parent compound is a band semiconductor \cite{Singh2009a, An2009}, there are no electronic states near $E_{\rm F}$. Thus, to visualize the location of the top of the valence band at the $\Gamma$ point $(k_x,k_y)=(0,0)$, we plot in Figs.~\ref{fig:Figure_ARPES}(a) and~2(b) the ARPES intensity map, at an energy 150~meV \emph{below} $E_{\rm F}$, and the energy distribution curves (EDCs), respectively, at 200~K for $x=0$.  Figure~\ref{fig:Figure_ARPES}(c) shows the intensity plot for $x=0.016$ \emph{at} $E_{\rm F}$ and the EDCs are shown in Fig.~\ref{fig:Figure_ARPES}(d) at 40~K, which demonstrate the presence of a finite-size Fermi surface at the $\Gamma$-point. This observation is consistent with the above measurements indicating metallic character for this doped composition.  The energy dispersions of the bands at the $\Gamma$ point for both samples in Figs.~\ref{fig:Figure_ARPES}(b) and~2(d) show that the dispersions are hole-like for  both $x = 0$ and~0.016, which is the same type of dispersion as for the hole band(s) at $\Gamma$ in BaFe$_2$As$_2$ \cite{Liu2008}. The Fermi surface in Fig.~\ref{fig:Figure_ARPES}(c) is shown in two high-symmetry directions, {\it i.e.} (0,0)--($\pi$,0) and (0,0)--($\pi$,$\pi$). Since the bottom of the electron band at the X point at ($\pi$,$\pi$) is above $E_{\rm F}$ (see Fig.~\ref{fig:Figure_BANDS} below), we do not expect to observe intensity from a bulk electron pocket at this position.  We return to this point below.

\begin{figure}
	\includegraphics[width=3in]{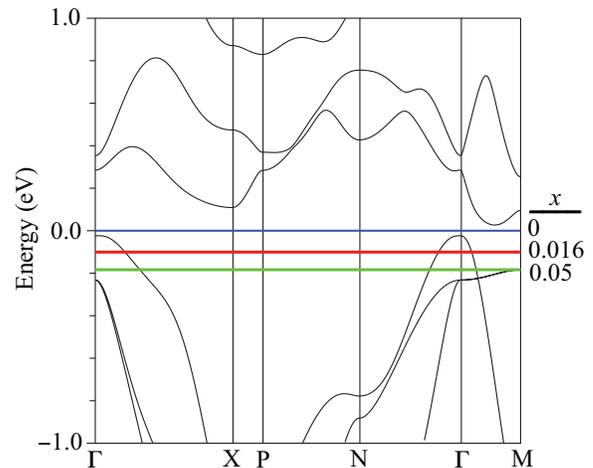}
	\caption{(Color online) Band structures of Ba$_{1-x}$K$_x$Mn$_2$As$_2$ with $x = 0,\ 0.016$, and 0.05 along high-symmetry directions of the Brillouin zone. The Fermi energies are indicated by horizontal lines for the hole concentrations labeled on the right-hand ordinate.  The labeling of the high-symmetry points is from Fig.~14 of Ref.~\cite{Johnston2010}.}
	\label{fig:Figure_BANDS}
\end{figure}

Bulk spin-polarized band structure calculations for the G-type AF structure are shown in Fig.~\ref{fig:Figure_BANDS} for $x=0$, 0.016 and 0.05.  The band gap for $x=0$ is 54.8~meV\@.  The bands do not change significantly with increasing $x$, but $E_{\rm F}$ decreases with increasing $x$ as shown in Fig.~\ref{fig:Figure_BANDS}, corresponding to rigid-band behavior.   The hole pocket Fermi surface at the $\Gamma$ point has a rounded square shape in the $ab$~plane whose maximal dimension increases with increasing $x$ as expected. The Fermi surface is closed and flattened along the $c$~axis (not shown), in contrast to the corrugated cylindrical hole Fermi surfaces at $\Gamma$ in the iron arsenides \cite{Johnston2010}.  The calculations also show that the bulk electron bands at the X point in Ba$_{1-x}$K$_x$Fe$_2$As$_2$ at $E_{\rm F}$ \cite{Liu2008} are not present in Ba$_{1-x}$K$_x$Mn$_2$As$_2$.  This is expected since Ba$_{1-x}$K$_x$Fe$_2$As$_2$ is a semimetal \cite{Johnston2010} whereas Ba$_{1-x}$K$_x$Mn$_2$As$_2$ is a hole-doped band semiconductor.  We have also carried out electronic structure calculations for the (001) surface states and have observed such states at $(\pi,0)$ and at X = $(\pi,\pi)$ at $E_{\rm F}$ for hole-doped compositions, consistent with the above ARPES measurements at these positions in $k$~space [Fig.~\ref{fig:Figure_ARPES}(c)].  Such surface states may also be present in the FeAs-based materials at the X-point but would be masked by the presence of the bulk electron pockets at the same position in $k$~space.

The calculated values of the Fermi wave vector $k_{\rm F}$ from the band structure calculations are $1.89\times10^9$ and $2.61\times10^9$~m$^{-1}$ for $x = 0.016$ and 0.05, respectively. These values match quite well the values calculated from the doping concentrations $n$ using the one-band result $k_{\rm F} = (3\pi^2n)^{1/3}$, where we get $k_{\rm F} = 1.59\times10^9$ and $2.33\times10^9$~m$^{-1}$ for the two compositions, respectively. The value of $k_{\rm F}$ estimated from the ARPES measurements for the single crystal with $x=0.016$ (Fig.~\ref{fig:Figure_ARPES}) is $0.1\pi/a-0.2\pi/a$ or, equivalently, $k_{\rm F} \approx 0.8-1.5\times10^9$~m$^{-1}$, which agrees with the above-calculated value for this $x$. The calculated bare band structure densities of states $N$ at $E_{\rm F}$ are 0.315 and 0.819 states/(eV\,f.u.) for both spin directions for $x=0.016$ and 0.05, respectively.  These values are significantly smaller than estimated above from the specific heat $\gamma$ values, suggesting the presence of many-body enhancement effects.  The orbital-decomposed $N(E_{\rm F})$ consists of 66\%~Mn 3$d$ and 28\% As 4$p$ for $x=0.016$ and 60\%~Mn 3$d$ and 30\% As 4$p$ for $x=0.05$.  These data suggest that the electronic conduction is mainly due to itinerant As 4$p$ holes, because the effective mass of these holes is expected to be much less than the effective mass of holes with Mn~$3d$ character.

\begin{figure}
	\includegraphics[width=2.5in]{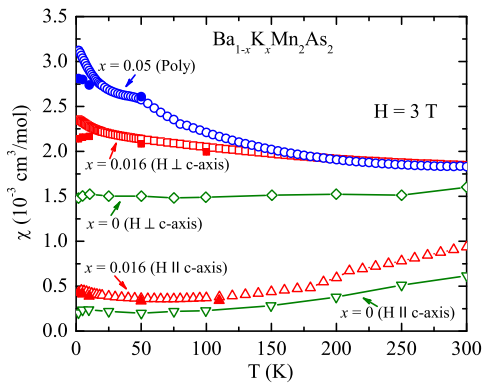}
	\caption{(Color online) Magnetic susceptibility $\chi \equiv M/H$ versus temperature $T$ of a crystal of ${\rm Ba_{0.984}K_{0.016}Mn_2As_2}$ and of polycrystalline ${\rm Ba_{0.95}K_{0.05}Mn_2As_2}$, where {\bf M} is the sample magnetization and {\bf H} is the applied magnetic field. For the crystal {\bf H} was either parallel to the $c$-axis ($\chi_c$) or in the $ab$~plane ($\chi_{ab}$).  The solid symbols at low $T$ are the intrinsic high-field slopes of $M(H)$ isotherms indicating the influence of saturable paramagnetic impurities. Data for single-crystalline BaMn$_2$As$_2$ \cite{Singh2009a} are shown for comparison.}
	\label{fig:Figure_MT}
\end{figure}

The anisotropic magnetic susceptibilities $\chi_{ab}$ and $\chi_{c}$ versus $T$ of a single crystal with $x=0.016$ and the $\chi$ of a polycrystalline sample with $x=0.05$ are shown in Fig.~\ref{fig:Figure_MT}. Data for an undoped $\rm{BaMn_2As_2}$ single crystal are shown for comparison \cite{Singh2009a}. The $\chi_{c}(T)$ for $x=0.016$ is qualitatively similar to that of $\rm{BaMn_2As_2}$. On the other hand, in contrast to the $\chi_{ab}(T)$ of $\rm{BaMn_2As_2}$ which remains nearly $T$ independent below 300~K, $\chi_{ab}(T)$ for $x=0.016$ increases with decreasing $T$ down to the lowest measurement $T$. These observations indicate that although the underlying AF structure has not changed and most probably is still a G-type or closely related structure, there are significant perturbations caused by K-doping. The dependence of the anisotropy versus $T$ and its similarity to that of undoped $\rm{BaMn_2As_2}$ suggest that $T_{\rm N}$ is significantly above 300~K\@. The $\chi(T)$ for the polycrystalline sample with $x=0.05$ monotonically increases with decreasing $T$ below 300~K and exhibits a kink at 50~K of unknown origin. The magnitude of $\chi(T)$ of this $x=0.05$ sample is significantly larger than that of the powder-averaged single crystal $\chi(T)$ data for either $x=0$ or 0.016.  This discrepancy again indicates that the emergence of metallic behavior upon K-doping leads to additional contributions and/or changes to the magnetism.  The $M(H)$ isotherms for this $x = 0.05$ sample showed no evidence for any ferromagnetic impurities at any temperature between 1.8 and 300~K\@.

\begin{figure}
	\includegraphics[width=2in]{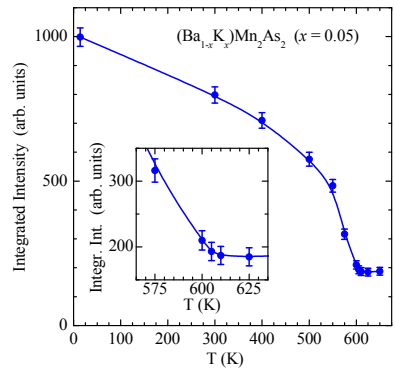}
	\caption{(Color online) Integrated intensity of the (1 0 1) magnetic Bragg peak measured for Ba$_{1-x}$K$_x$Mn$_2$As$_2$ ($x=0.05$).  The solid line is a guide to the eye. Note that there is a low intensity nuclear Bragg peak at (1 0 1) present at all temperatures.  Inset: Expanded plot near $T_{\rm N}=607(2)$~K\@.}
	\label{fig:Figure_ND}
\end{figure}

From Rietveld refinement of our ND data for the polycrystalline sample with $x=0.05$, the chemical unit cell is well-described at 14~K by the bct ${\rm ThCr_2Si_2}$ structure (space group $I4/mmm$) with lattice parameters $a=b= 4.1566(6)$~\AA, $c= 13.4043(2)$~\AA, and As position parameter $z_{\rm As}$ = 0.3642(5).  As was found for the $\rm{BaMn_2As_2}$ ($x=0$) parent compound \cite{Singh2009b}, the K-doped system remains in the bct phase at all measured temperatures. The ND data for $x=0.05$ at 14~K indicate a slight contraction of the $c$~axis lattice parameter with respect to the parent system at 10~K [$a = b = 4.1539(2)$~\AA, $c = 13.4149(8)$~\AA] and a corresponding 0.8\% increase in the As position parameter [$z_{\rm As}$ = 0.3613(2)] \cite{Singh2009b}. The modeling of the magnetic scattering revealed a G-type AF spin structure below $T_{\rm N} = 607(2)$~K (Fig.~\ref{fig:Figure_ND}).  The ordered moment at $T = 14$~K is $\mu=4.21(6)~\mu_{\rm B}$/Mn aligned along the $c$~axis. The G-type AF order found for $x = 0.05$ is identical to that previously determined for the parent $\rm{BaMn_2As_2}$ compound, which gave $T_{\rm N}=625(1)$~K and $\mu=3.88(4)~\mu_{\rm B}$/Mn at $T = 10$~K \cite{Singh2009b}. Our refined value of $\mu$ for $x = 0.05$ is slightly larger than that of the parent compound, which may suggest a reduction in the spin-dependent hybridization between the Mn $3d$ states and the As $4p$ states \cite{An2009}.

In conclusion, we succeeded in doping the antiferromagnetic insulator $\rm{BaMn_2As_2}$ into a hole-doped metallic state by partial substitution of Ba by K to form the new system Ba$_{1-x}$K$_x$Mn$_2$As$_2$.  Both single-crystalline ($x=0.016)$ and polycrystalline ($x=0.05$) samples were synthesized and studied.  The $\rho(T)$, $C_{\rm p}(T)$ and ARPES data and our electronic structure calculations consistently indicate metallic character for these compositions.  Our ND and $\chi$ measurements demonstrate that the local moment magnetism of undoped ${\rm BaMn_2As_2}$ is preserved in the metallic doped samples. Thus the Ba$_{1-x}$K$_x$Mn$_2$As$_2$ system is a bridge between the layered high-$T_{\rm c}$ cuprates and iron arsenides.  Our results on the metallic MnAs-based materials suggest many opportunities for theoretical and experimental investigations examining the relationships between the layered transition metal arsenide and cuprate classes of materials and offer the potential for new superconductors.  The roles of the local Mn magnetic moments and their AF interactions in determining $T_{\rm c}$ would be particularly interesting to study.

The work at Ames Laboratory was supported by the Department of Energy-Basic Energy Sciences under Contract No.~DE-AC02-07CH11358.

\emph{Note added in proof}.---ÑAdditional papers related to our work have recently appeared \cite{Satya2011, Simonson2011, Sun2011, Bau2012}.

\end{document}